\documentclass[preprintnumbers, superscriptaddress, showpacs, nofootinbib, amsfonts, amsmath, amssymb, aps, prd, notitlepage, floatfix]{revtex4-2}

\usepackage{graphicx}
\usepackage{curves}
\usepackage{hyperref}
\usepackage{amsmath}
\usepackage{graphicx}
\usepackage{bbold}
\usepackage{amssymb}
\usepackage{mathtools}
\usepackage{physics}
\usepackage{slashed}
\usepackage{xcolor}
\usepackage{MnSymbol}

\usepackage{subcaption}
\usepackage{lipsum}
\usepackage{amssymb}

\usepackage{booktabs}
\usepackage{siunitx}
\usepackage{multirow}
\usepackage{orcidlink}

\usepackage{lipsum}
\begin{document}

\title{Extrapolation to infinite model space of no-core shell model calculations using machine learning}

\author{Aleksandr Mazur\orcidlink{0000-0001-9230-9256}}
\email[Email: ]{000287@togudv.ru}
\affiliation{Laboratory for Modeling Quantum Processes, Pacific National University, 136 Tikhookeanskaya Street\\
Khabarovsk, 680035, Russia}

\author{Roman Sharypov\orcidlink{0000-0002-6324-7217}}
\email[Email: ]{2017104939@togudv.ru}
\affiliation{Laboratory for Modeling Quantum Processes, Pacific National University, 136 Tikhookeanskaya Street\\
Khabarovsk, 680035, Russia}

\author{Andrey Shirokov\orcidlink{0000-0002-0331-6209}}
\email[Email: ]{shirokov@nucl-th.sinp.msu.ru}
\affiliation{Skobeltsyn Institute of Nuclear Physics, Lomonosov Moscow State University, Leninskie Gory 1/2\\
Moscow, 119991, Russia}

\begin{abstract}
An ensemble of neural networks is employed to extrapolate no-core shell model (NCSM) results to infinite model space for light nuclei. We present a review of our neural network extrapolations of the NCSM results obtained with the Daejeon16 $NN$ interaction in different model spaces and with different values of the NCSM basis parameter $\hbar\Omega$  for energies of nuclear states and root-mean-square (rms) radii of proton, neutron and matter distributions in light nuclei. The method yields convergent predictions with quantifiable uncertainties. Ground-state energies for $^{6}$Li, $^{6}$He, and the unbound $^{6}$Be, as well as the excited $(3^{+},0)$ and $(0^{+},1)$ states of $^{6}$Li, are obtained within a few hundred keV of experiment. The extrapolated radii of bound states converge well. In contrast, radii of unbound states in $^{6}$Be and $^{6}$Li do not stabilize.

These results demonstrate that machine-learning extrapolations can extend the reach of \textit{ab initio} nuclear structure calculations with reliable accuracy.
\end{abstract}

\keywords{No-core shell model; 
\textit{ab initio} approaches;
extrapolation of variational calculations;
machine learning;
artificial neural networks.}

%%%%%%%%%%%%%%%%%%%%%%%%%%%%%%%%%%%%%%%%%%%%%%%%%%%%%%
%%%%%%%%%%%%%%%%%%%%%%%%%%%%%%%%%%%%%%%%%%%%%%%%%%%%%%
%%%%%%%%%%%%%%%%%%%%%%%%%%%%%%%%%%%%%%%%%%%%%%%%%%%%%%

\maketitle

%% main text

\section{Introduction}
\label{introduction}

One of the most promising {\it ab initio} approaches for the study of light atomic nuclei---approaches that do not rely on model assumptions or phenomenological approximations---is the No-Core Shell Model (NCSM)~\cite{Barrett_2013}. In the NCSM all nucleons are treated as spectroscopically active, and the only input that determines the properties of the nucleus under investigation is a realistic $NN$ (nucleon-nucleon) interaction. The many-body basis states in the NCSM are Slater determinants constructed from single-particle oscillator functions and include all possible nucleon configurations with total excitation quanta up to $N_{\max}$. The parameters of the NCSM are the oscillator energy $\hbar\Omega$ and the truncation parameter $N_{\max}$ that define the model space.

Results of the NCSM are exact in the limit $N_{\max}\rightarrow\infty$. However, the dimension of the many-body basis grows exponentially with $N_{\max}$, and even on modern supercomputers, NCSM calculations with reasonable accuracy can be performed only for nuclei with mass numbers $A\lesssim20$. Therefore, the improvement of existing methods and the development of new extrapolation methods that extend NCSM and other {\it ab initio} results to effectively infinite model spaces remain an important and timely problem.

Most contemporary extrapolation techniques are based on exponential or polynomial fits to results obtained in small model spaces and provide reasonable approximations to bound-state energies~\cite{Bogner2008, Roth_2009, Maris2009, Jurgenson_2013, Furnstahl_2012, Coon_2012, More_2013, Furnstahl_2014, Furnstahl_2015, Wendt_2015}. These methods have been generalized~\cite{Furnstahl_2012, Coon_2012, More_2013, Furnstahl_2014, Furnstahl_2015, Wendt_2015, Rodkin_2022} to permit extrapolation of observables that do not obey a variational principle. For example, Ref.~\cite{Rodkin_2022} proposed the Twisted Tape Extrapolation (TTE) method, which is based on an analysis of the two-dimensional dependence of results on $\hbar\Omega$ and $N_{\max}$. This method was successfully applied to the study of the root-mean-square (rms) radii of the point-proton ($r_p$), point-neutron ($r_n$) and point-nucleon (matter) ($r_m$) distributions in $^6$He~\cite{Rodkin_2022} and $^6$Li~\cite{Rodkin2023}. All of the above techniques are essentially phenomenological and lack a rigorous theoretical foundation, which limits their range of applicability.

A physically motivated alternative is the approach of Ref.~\cite{Shirokov_2021}, in which the extrapolated bound-state energy is determined by locating the poles of the $S$-matrix.

The use of machine-learning methods---which are increasingly employed in both theoretical and experimental nuclear physics (see the review~\cite{Boehnlein_2022})---is a promising direction for addressing the extrapolation problem.

The possibility of using ensembles of artificial neural networks to extrapolate NCSM results to large model spaces was first investigated in Refs.~\cite{Negoita_2018, Negoita_2019}. In the proposed approach (below referred to as ISU) simple feed-forward neural networks with a single hidden layer were employed; these networks were trained on a collection of NCSM calculations of the ground-state energy $E$ and the point-proton
%root-mean-square
rms
radius $r_{p}$ for a given nucleus across different model spaces $N_{\max}$ and sets of $\hbar\Omega$ values. Similar network-based extrapolation strategies for NCSM results were also used in Refs.~\cite{Jiang_2019, Vidana_2023}.

 An alternative strategy was developed by the TUDa group~\cite{Knoll_2023, Wolfgruber_2024}. In this approach, ensembles of universal neural networks are constructed for each observable using NCSM calculations that converge to exact values for the lightest nuclei $^2$H, $^3$H and $^4$He with various $NN$ interactions; these pretrained ensembles are then applied to directly extrapolate results obtained in smaller NCSM model spaces for heavier nuclei. A comparison of the ISU and TUDa extrapolations for ground-state energies $E$ and point-proton rms radii $r_p$ of Li isotopes demonstrates consistency between the two approaches~\cite{Knoll_2025}.

In Refs.~\cite{Mazur_2024, Sharypov_2024} we continued the idea of the ISU approach~\cite{Negoita_2018, Negoita_2019}. We proposed a more sophisticated neural network topology, investigated the specifics of constructing the training dataset, and formulated strict criteria for selecting trained networks.
As a result, we obtained a universal extrapolation method applicable to {\it ab initio} calculations
%performed in small model spaces; 
this method mitigates overfitting of the network, manages the need to split input data into separate training and test sets, and provides high statistical reliability of the extrapolated results.

In the present contribution, we discuss the developed approach to the investigation the properties of the ground states of $^6$He ($J^\pi, T = 0^+,1$) and $^6$Li ($1^+,0$). We also examine characteristics of the unbound ground state of  $^6$Be ($0^+,1$) and of the excited states, bound ($3^+,0$) and unbound ($0^+,1$), of $^6$Li. We utilize NCSM results obtained with the Daejeon16 $NN$ interaction~\cite{Shirokov_2016} using  MFDn code \cite{Aktulga_2014, Maris_2010, Maris_2022, Cook_2022} in model spaces with $N_{\max}=4,6,8,\dots,18$, we perform extrapolations of the energies $E$ and the rms point-radii $r_p$, $r_n$ and $r_m$ for the states listed above.

We analyze the convergence of the extrapolated results as data from progressively larger model spaces are included in the training set.
%allows a detailed investigation of their properties.

\section{Method}
\label{Method}

We employ an ensemble approach based on fully connected feedforward neural networks to extrapolate observables obtained from the NCSM calculations to large model spaces $N_{\max}$, where the NCSM results are expected to be converged. The ensemble both improves predictive accuracy and provides an empirical uncertainty estimate. The method was validated~\cite{Mazur_2024, Sharypov_2024} on the deuteron ground-state energy computed with the realistic Nijmegen~II $NN$ potential~\cite{PhysRevC.49.2950}, a case notable for
%slow and $N_{\max}$ parity-dependent convergence
odd-even staggering of the results with increasing $N_{\max}$
and nontrivial $\hbar\Omega$ dependence. The method of extrapolation employed is described in most detail in Ref. \cite{Sharypov_2024}.

%\subsection{Neural network design}
%\label{Method: Neural network design}
The neural network ensemble comprises 1024 identically structured multilayer perceptrons. Each network receives two inputs, $N_{\max}$ and $\hbar\Omega$, and has the layer layout $2$--$10$--$10$--$10$--$1$ where 
each number is the number of neurons in the successive layers.
%numbers indicate neuron quantity per layer.
The first hidden layer uses the identity activation function, the second and third hidden layers use the sigmoid activation function, and the output layer also uses the identity activation function. A consequence of this network configuration is the convergence of predictions as $N_{\max}$ increases. 
Weight initialization follows the Glorot Uniform~\cite{Glorot_pmlr-v9-glorot10a} scheme. We implement the models in Keras~\cite{chollet2015keras} with TensorFlow backend~\cite{tensorflow2015-whitepaper}.

The network inputs are $N_{\max}$ and $\hbar\Omega$, while the outputs are computed observables (energies, radii, etc.). Following the selection rules of Ref.~\cite{Shirokov_2021}, for the training we retain, for each model space, data with $\hbar\Omega$ values to the right of the variational minimum for extrapolating energies; for the radius extrapolations we apply practical cutoffs ($12.5 \leq  \hbar\Omega\le 40\,$ MeV).  All data are scaled to the unit interval $[0,1]$ prior to training.

Training is conducted with the Adam~\cite{Adam} optimizer with a triangular cyclical learning rate schedule~\cite{smith2017cyclical} implemented via TensorFlow Addons~\cite{TensorFlowAddons}; the learning rate cycles between $10^{-4}$ (base) and $10^{-2}$ (initial amplitude) with a decaying amplitude. The minimized loss function is the mean squared error; the batch size equals the training set length. Networks are trained for $10^{6}$ epochs (see Refs.~\cite{Mazur_2024}, ~\cite{Sharypov_2024} for further implementation details).

%\subsection{Prediction processing}
%\label{Method: prediction processing}

For each trained network, we generate predictions for model spaces $N_{\max}$ up to a large final value $N_{\max}^{f}$ (we use $N_{\max}^{f}=300$) over the same $\hbar\Omega$ range as in training. Not all trained networks yield physically consistent extrapolations; therefore, a multi-stage filtering is applied to form the final ensemble. The filtering process described below is utilized both for energies and radii.

Networks are retained only if they satisfy (i) a ``soft'' variational principle for energies, (ii) $\hbar\Omega$-stability at $N_{\max}^{f}$, and (iii) their predictions converge sufficiently quickly. From networks passing these criteria, we additionally discard 5\% with the largest final loss function value.

%\paragraph{Outlier removal and final estimate.} 
For the surviving networks we extract, at $N_{\max}^{f}$, either the minimal value over $\hbar\Omega$ (for energies) or the mean over $\hbar\Omega$ (for radii), producing an ensemble distribution of extrapolated values. Outliers are removed using a boxplot rule~\cite{wilcox2009basic}. The median of the filtered distribution is taken as the final estimate, and the interquartile range (25th--75th percentile) is utilized as the method uncertainty.

All hyperparameters and selection thresholds were chosen and tested on
%the described deuteron benchmark
various light nuclei; further implementation particulars and numerical experiments are provided in Refs.~\cite{Mazur_2024, Sharypov_2024}.

\section{Results}
\label{Results}
Based on the NCSM calculations with the Daejeon16 $NN$ interaction \cite{Shirokov_2016}, we performed extrapolations of energies $E$ and rms  point-proton, point-neutron, and matter radii ($r_p$, $r_n$, $r_m$) for the ground states of $^{6}$He, $^{6}$Li, $^{6}$Be and for the excited states $(0^{+},1)$ and $(3^{+},0)$ of $^{6}$Li. The NCSM calculations were performed with a $\hbar\Omega$ step of 2.5 MeV. We note that increasing the number of training data by using a finer $\hbar\Omega$ grid does not sufficiently improve the precision of the extrapolated results.

Of particular interest for demonstrating our method \cite{Mazur_2024, Sharypov_2024} is the ground state of $^{6}$Li, which has been investigated with other neural-network models \cite{Negoita_2019, Negoita_2018, Knoll_2023, Wolfgruber_2024, Knoll_2025} based on the same set of NCSM calculations with the Daejeon16 $NN$ interaction in model spaces $4\le N_{\max}\le 18$ but using different training-data selections. In the ISU approach the training sets comprise data with $\hbar\Omega\in[8,50]\ \mathrm{MeV}$ (ISU-a~\cite{Negoita_2019, Negoita_2018}) or $\hbar\Omega\in[10,30]\ \mathrm{MeV}$ (ISU-b \cite{Knoll_2025}), whereas the TUDa approach \cite{Knoll_2023,Wolfgruber_2024,Knoll_2025} uses $\hbar\Omega\in[10,20]\ \mathrm{MeV}$.

\begin{figure}[th]
\centerline{
\includegraphics[width=6.9in]{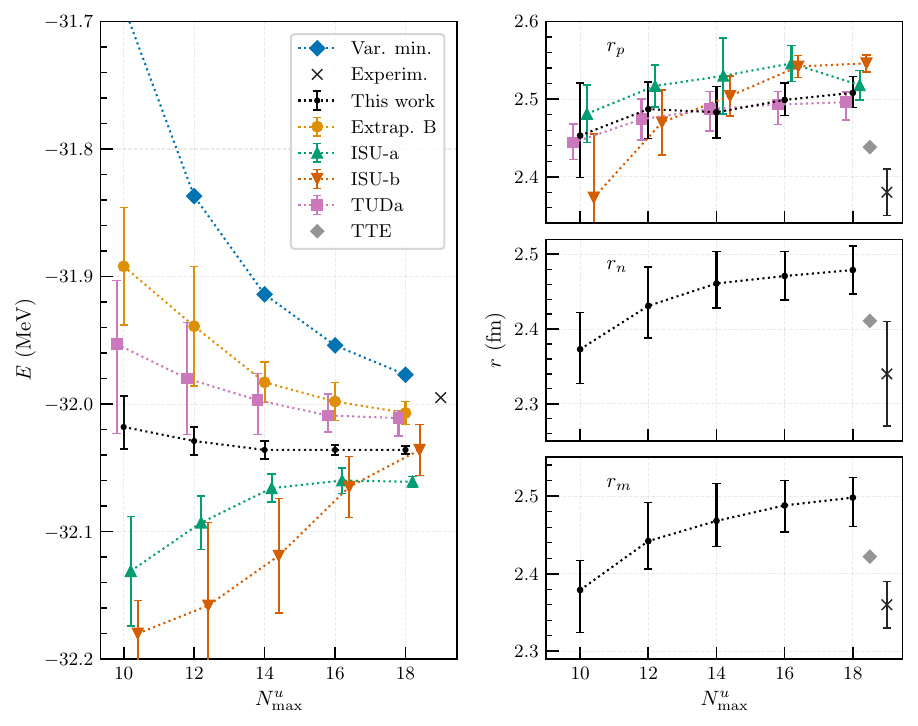}
}

% список изменений:
% * цветовая схема colorblind-friendly
% * радиусы курсивом и не жирным как в тексте
% * в радиусах небольшое изменение вертикальной оси, чтобы не выходили усы за рамки графика
% исправлено несоответсвие в r_p графика и таблицы: 2.411 TTE и 2.38+-0.03 exp
\caption{Convergence of the extrapolated ground-state energy (left) and rms radii (right) of $^{6}$Li. Symbols are explained in the legend.}
\label{Fig 6Li gs}
%\alttext{Extrapolation results for $^6$Li}
\end{figure}

All machine-learning based approaches produce ground-state energies of $^{6}$Li that lie below the widely used phenomenological Extrapolation~B \cite{Maris2009} (see the left panel of Fig.~\ref{Fig 6Li gs}). Compared with ISU and TUDa, our method demonstrates faster convergence when results from larger model spaces (up to $N^u_{\max}$) are included in the training set, and it yields smaller extrapolation uncertainties at all $N^u_{\max}$ (below $0.005\%$ for $N^u_{\max}=18$).

Importantly, the energy predictions for the $^{6}$Li ground state obtained with these different approaches are statistically distinguishable: only for $N^u_{\max}=10$ and $12$ our predictions overlap with TUDa within the stated uncertainties; otherwise, no such overlap is observed. We also note that TUDa predictions, including uncertainties, overlap with Extrapolation~B. The comparison of the two ISU versions illustrates the sensitivity of the extrapolated results to the construction of the training set: in the ISU-b case, the convergence is absent. Hence, the application of machine learning methods to extrapolation of nuclear observables requires careful preparation of the training dataset.

The first excited state $(3^{+},0)$ with experimental excitation energy of $E^{\mathrm{ex}}=2.186\ \mathrm{MeV}$ lies in the continuum and decays via the ${\alpha+d}$ channel with threshold at $0.712\ \mathrm{MeV}$ \cite{Tilley_2002}.
Interestingly, the decay of the higher second excited state $(0^{+},1)$ with $E^{\mathrm{ex}}=3.563\ \mathrm{MeV}$ \cite{Tilley_2002} into ${\alpha+d}$ is forbidden both by the spin-parity and by the isospin, while the next decay channel ${\alpha+n+p}$ has a threshold $136\ \mathrm{keV}$ higher in energy and is closed.
Therefore this is a bound state and its width of 
%the $(0^{+},1)$ state, 
$\Gamma=8.2\ \mathrm{eV}$, is determined solely by its electromagnetic decay to the ground state and is negligible compared to the width of the $(3^{+},0)$ state, $\Gamma=20\ \mathrm{keV}$.

\begin{figure}[th]
\centerline{
\includegraphics[width=6.9in]{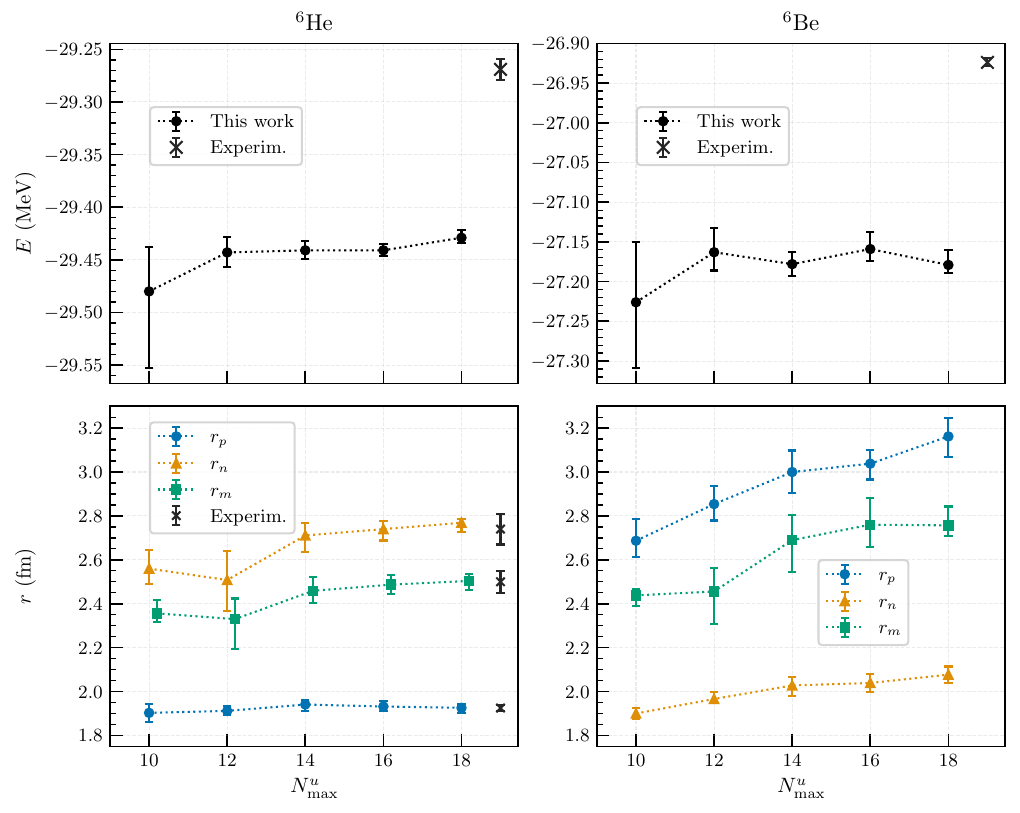}
}
\caption{
Convergence of the extrapolated energies (top) and rms radii (bottom) 
for the ground states 
$^{6}$He (left) and $^{6}$Be (right) nuclei. 
%Experimental data are shown by open symbols.%
}
\label{fig: 6He 6Be}
%\alttext{Extrapolation results for $^6$He and  $^6$6Be}
\end{figure}

\begin{figure}[th]
\centerline{
\includegraphics[width=6.9in]{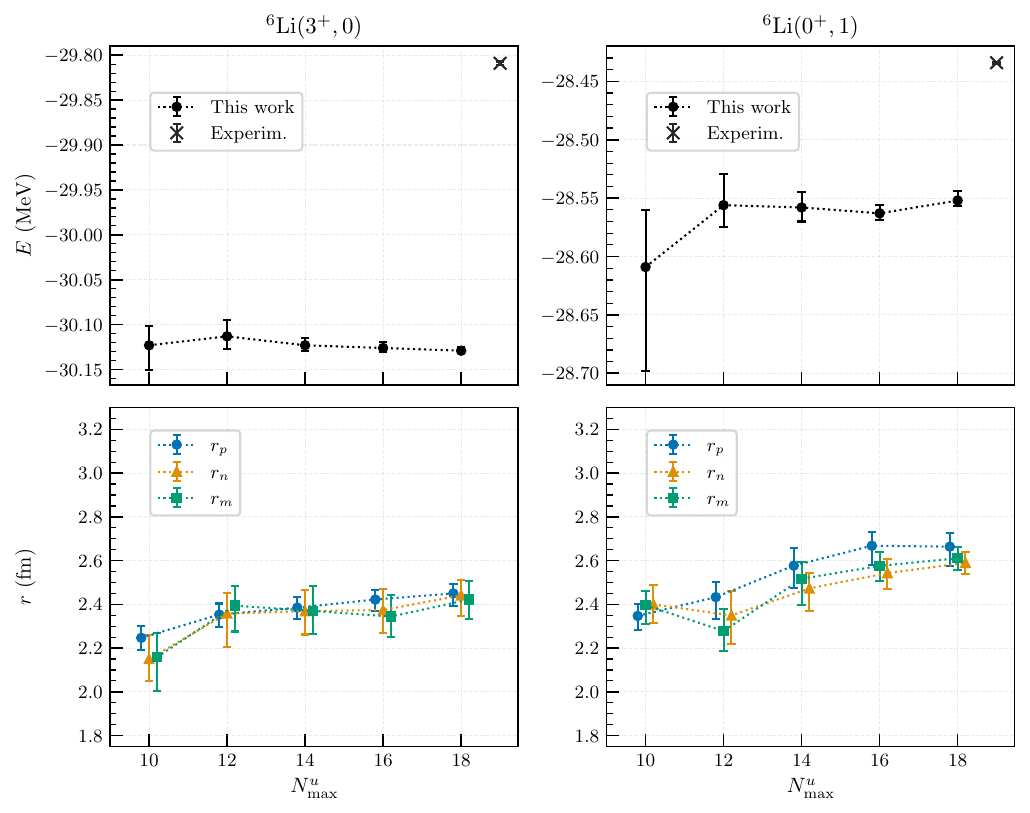}
}
\caption{
Convergence of the extrapolated energies (top) and rms radii (bottom) 
for the excited states of $^{6}$Li($3^{+},0$) (left), $^{6}$Li($0^{+},1$) (right). 
%Experimental data are shown by open symbols.%
}
\label{fig 6Li excited}
%\alttext{Extrapolation results for $^6$Li 0+ 1, 3+ 0}
\end{figure}

The second excited state $(0^{+},1)$ of $^{6}$Li and the ground states of $^{6}$He and $^{6}$Be form an isospin triplet. The ground state of $^{6}$He is weakly bound. The proton-rich ground state of $^{6}$Be is unbound and decays through the ${\alpha+p+p}$ channel, forming a resonance with the energy of $1.371\ \mathrm{MeV}$ and the width of $92\ \mathrm{keV}$. Despite these substantial physical differences, the energy extrapolations (see Figs.~\ref{fig: 6He 6Be}, \ref{fig 6Li excited}) exhibit good convergence in all cases: predicted values stabilize starting from $N^{u}_{\max}=12$ and their uncertainties decrease. 
This behavior also holds for the decaying unbound $(3^{+},0)$ state of $^{6}$Li. Thus, machine-learning methods can be used to extrapolate the energies of at least for relatively narrow nuclear resonances with widths up to $100\ \mathrm{keV}$.

Final energy predictions for the ground states of $^{6}$Li, $^{6}$He, $^{6}$Be and for the excited states $(0^{+},1)$ and $(3^{+},0)$ of $^{6}$Li based on training sets with $N_{\max}^u=18$, together with experimental data and results of other neural network extrapolations and of phenomenological Extrapolation~B, are reported in Table~\ref{tab energy}. The largest relative energy uncertainty at $N^{u}_{\max}=18$ is observed for $^{6}$Be and does not exceed approximately $0.01\%$. 

\begin{table}[pt]
\caption{Extrapolated energies (in MeV) of selected states of $^{6}$He, $^{6}$Li, and $^{6}$Be obtained with $N_{\max}$ up to 18.
Prediction uncertainties are given in parentheses for symmetric errors; for asymmetric distributions upper and lower limits are indicated. Experimental data are taken from Ref.~\cite{Tilley_2002}.}
\label{tab energy}
\centering
\begingroup
\renewcommand{\arraystretch}{1.40} % increase row spacing
\begin{ruledtabular}
\begin{tabular}{@{} l  l  l  l @{}}
\toprule
Nucleus & Our result & Experiment & Other extrapolations \\
\midrule
$^{6}$He (g.s.) &
  $-29.429^{+0.007}_{-0.005}$ &
  $-29.269(10)$ &
  $-29.41(1)$ (Extrap.\ B) \\[6pt]

\multirow{3}{*}{$^{6}$Li (g.s.)} 
  & \multirow{3}{*}{$-32.036(3)$} 
  & \multirow{3}{*}{$-31.995$} 
  & $-32.007(9)$ (Extrap.\ B) \\ 
  & & & $-32.061(4)$ (ISU-a) \\ 
  & & & $-32.011^{+0.006}_{-0.014}$ (TUDa) \\[6pt]

$^{6}$Li ($3^{+},0$) &
  $-30.129^{+0.004}_{-0.003}$ &
  $-29.809(2)$ &
  $-30.10(1)$ (Extrap.\ B) \\[6pt]

$^{6}$Li ($0^{+},1$) &
  $-28.552^{+0.008}_{-0.005}$ &
  $-28.434(1)$ &
  $-28.507(4)$ (Extrap.\ B) \\[6pt]

$^{6}$Be (g.s.) &
  $-27.18^{+0.02}_{-0.01}$ &
  $-26.924(5)$ &
  $-27.13(2)$ (Extrap.\ B) \\
\bottomrule
\end{tabular}
\end{ruledtabular}
\endgroup

\vspace{2mm}
\raggedright
\footnotesize
Central values are given in the numeric columns. Numbers in parentheses denote uncertainties; asymmetric uncertainties are given as superscript/subscript.
\end{table}
\renewcommand{\arraystretch}{1.0}

Careful extrapolations based on machine learning indicate that the Daejeon16 interaction produces modest overbindings of the considered states by $41\ \mathrm{keV}$, $160\ \mathrm{keV}$, $256\ \mathrm{keV}$, $320\ \mathrm{keV}$ and $118\ \mathrm{keV}$ for the ground-state energies of $^{6}$Li, $^{6}$He, $^{6}$Be and for the $(3^{+},0)$ and $(0^{+},1)$ states of $^{6}$Li, respectively. The predicted excitation energies of the $(3^{+},0)$ and $(0^{+},1)$ states of $^{6}$Li (1.907~MeV and 3.484~MeV) are slightly smaller than the respective experimental values (2.186~MeV and 3.563~MeV \cite{Tilley_2002}).

Convergence of the rms point-radii for the $^{6}$Li ground state is slower and the relative prediction uncertainties are larger (of order $1\%$ at $N^u_{\max}=18$; see the right panel of Fig.~\ref{Fig 6Li gs}). The final predictions for $r_p$, $r_n$ and $r_m$ in the $^{6}$Li ground state (Table~\ref{tab radii}) are 
%consistent with one another 
the same
within the uncertainties, reflecting the equal number of protons and neutrons and agreeing qualitatively with the experiment. However, our predictions overestimate the experimental radii \cite{Tanihata2013} and are larger than the TTE extrapolation results \cite{Rodkin2023}. At the same time, our predictions for $r_p$ are in good agreement with the TUDa and ISU-a approaches.

\begin{table*}[t]
\caption{Results of extrapolations of root-mean-square radii (in~fm) obtained with $N_{\max}^u=18$.
Experimental data are taken from the compilation~\cite{Tanihata2013}.
TTE extrapolation results are from~\cite{Rodkin_2022} ($^{6}$He) and~\cite{Rodkin2023} ($^{6}$Li).}
\label{tab radii}
\centering
\begingroup
\renewcommand{\arraystretch}{1.40} % increase row spacing for readability
\begin{ruledtabular}
\begin{tabular}{@{} l c l l l @{}}
Nucleus & Observable & Our result & Experiment & Other extrapolations \\
\hline
\multirow{3}{*}{$^{6}\mathrm{He}$ (g.s.)}
  & $r_{p}$ & 1.930(2)                        & 1.925(12) & 1.871(6) (TTE) \\
  & $r_{n}$ & $2.770^{+0.02}_{-0.04}$         & 2.74(7)  & 2.663(3) (TTE) \\
  & $r_{m}$ & $2.500^{+0.03}_{-0.04}$         & 2.50(5)  & 2.430(6) (TTE) \\
\hline
\addlinespace[8pt]
\multirow{5}{*}{$^{6}\mathrm{Li}$ (g.s.)}
  & \multirow{3}{*}{$r_{p}$}
    & 2.51(2)                        & 2.38(3)  & \begin{tabular}[t]{@{}l@{}}
                                                          2.438 (TTE)\\
                                                          2.518(19) (ISU-a)\\
                                                          $2.496^{+0.013}_{-0.023}$ (TUDa)
                                                      \end{tabular} \\
  & $r_{n}$ & 2.48(3)                        & 2.34(7)   & 2.411 (TTE) \\
  & $r_{m}$ & $2.500^{+0.03}_{-0.04}$         & 2.36(3)   & 2.422 (TTE) \\

\addlinespace[8pt]
\hline
\multirow{3}{*}{$^{6}\mathrm{Li}\ (0^{+},1)$}
  & $r_{p}$ & $2.660^{+0.06}_{-0.09}$         & \multicolumn{1}{c}{--} & 2.502 (TTE) \\
  & $r_{n}$ & 2.590(5)                        & \multicolumn{1}{c}{--} & 2.443 (TTE) \\
  & $r_{m}$ & 2.610(5)                        & \multicolumn{1}{c}{--} & 2.467 (TTE) \\
\end{tabular}
\end{ruledtabular}
\endgroup
\vspace{2mm}
\raggedright
\footnotesize Central values are given in the numeric columns. Numbers in parentheses denote uncertainties; asymmetric uncertainties are given as superscript/subscript.
\end{table*}
\renewcommand{\arraystretch}{1.0}

Convergence of the extrapolated $r_p$, $r_n$ and $r_m$ for the weakly-bound ground state of $^{6}$He (see Fig.~\ref{fig: 6He 6Be}) is no worse than for the $^{6}$Li ground state. The predicted values and uncertainties stabilize starting from $N^u_{\max}=14$ and agree well with the experiment. The proton radius $r_p$ of $^{6}$He is significantly smaller than $r_n$
%owing 
due
to the presence of a pair of weakly-bound neutrons. The uncertainties of the radius extrapolations for $^{6}$He are similar to those for the $^{6}$Li ground state; notably, within the uncertainties the predicted matter radius $r_m$ of $^{6}$He coincides with the $r_p$, $r_n$ and $r_m$ values in the $^{6}$Li ground state.

For the $(0^{+},1)$ state of $^{6}$Li, which is the bound isospin analogue of the $^{6}$He ground state, we observe similar convergence trends for the rms radii. As for the $^{6}$Li ground state, the extrapolated values of $r_p$, $r_n$ and $r_m$ agree within uncertainties, but both the extrapolated radii themselves and their uncertainty estimates for the $(0^{+},1)$ excited state are somewhat larger than that in the $^{6}$Li ground state. The final predictions for the radii of the $^{6}$Li and $^{6}$He ground states and for the $(0^{+},1)$ state of $^{6}$Li are slightly larger than the TTE results (see Table~\ref{tab radii}).

The third member of the isospin triplet, the proton-rich ground state of $^{6}$Be, is unbound and decays via ${\alpha+p+p}$. As shown in Fig.~\ref{fig: 6He 6Be}, the proton radius $r_p$ increases with $N^u_{\max}$, i.e. no convergence of $r_p$ is observed, which is expected for a resonant decaying state. The neutron radius $r_n$ of this nucleus is substantially smaller than the proton radius and also increases, albeit more slowly. The matter radius $r_m$ takes intermediate values and likewise does not display convergence.

For the unbound $(3^{+},0)$ state of $^{6}$Li, one would expect lack of convergence for the extrapolated radii $r_p$, $r_n$ and $r_m$. Nevertheless, as seen in Fig.~\ref{fig 6Li excited}, the extrapolations of these radii stabilize in the region $N^u_{\max}=12\text{--}16$, and only at $N^u_{\max}=18$ is a small increase of the extrapolated values is observed, possibly indicating the onset of divergence upon further enlargement of the training set by inclusion of results obtained with larger $N_{\max}$. This behaviour is plausibly an artifact of the slow convergence of the deuteron binding energy: in the NCSM calculations with $N_{\max}\le 16$ and various $\hbar\Omega$ values, most of the results correspond to a situation in which the $(3^{+},0)$ state appears bound and the ${\alpha+d}$ decay channel is closed due to the insufficient deuteron binding energy. At the same time, some training-data points corresponding to the optimal $\hbar\Omega$ interval for the deuteron convergence open the decay channel, and the presence of such heterogeneous data in the training set increases the extrapolation uncertainties for the radii of this state. It is to be expected that the NCSM calculations at larger $N_{\max}$ will indeed produce divergent extrapolations of the radii for this state.

\section{Discussion}
\label{Discussion}
In the machine learning approach proposed in Ref.~\cite{Negoita_2019} and improved by us in Refs.~\cite{Mazur_2024, Sharypov_2024}, we performed extrapolations of energies and rms point-proton, point-neutron, and point-nucleon (matter) radii. We utilize the NCSM results obtained with the $NN$ interaction Daejeon16 using MFDn code \cite{Aktulga_2014, Maris_2010, Maris_2022, Cook_2022} in model spaces up to $N^u_{\max}=18$. Also, this approach has been applied to the analysis of unbound resonant states.  Thanks to the use of a large ensemble of artificial neural networks with the proposed topology, careful construction of the training set, and strict selection criteria for trained networks, our method promises high reliability of the predictions.
%For the first time, this method was applied to the  electric quadrupole (E2) observables in $^{10}$Be and $^{10}$C nuclei. 

These findings highlight machine learning extrapolations as a robust new tool for nuclear theory.
We demonstrate that machine learning methods are applicable to extrapolation of energies not only for bound states but also for resonant states in nuclei, at least for resonances with widths up to 100~keV. The extrapolated energies of the ground states of $^{6}$He, $^{6}$Li, $^{6}$Be and of the excited states $(3^{+},0)$ and $(0^{+},1)$ of $^{6}$Li obtained in NCSM calculations with the $NN$ interaction Daejeon16~\cite{Shirokov_2016} lie below the results of the phenomenological Extrapolation~B \cite{Maris2009} and below the experimental values --- by 41~keV for the ground state of $^{6}$Li up to 320~keV for the $(3^{+},0)$ excited state of $^{6}$Li. The energy extrapolations for both bound and unbound states show good convergence with increasing $N^u_{\max}$, with relative prediction errors not exceeding 0.01\%.

Unlike energies, convergence of the extrapolated rms point radii is observed only for bound states: the ground states of $^{6}$Li and $^{6}$He and the $(0^{+},1)$ state of $^{6}$Li (with relative uncertainties of order ${\sim}1\%$). The predicted point radii $r_p$, $r_n$, and $r_m$ of the ground state of $^{6}$Li are larger than the experimental values, whereas for the weakly bound ground state of $^{6}$He the extrapolated radii agree with the experiment within the stated uncertainties.

For the extrapolation of the rms point radii of the unbound ground state of the proton-rich $^{6}$Be no convergence is observed: the predicted radii increase with $N^u_{\max}$. It is likely that, even if results from the NCSM calculations with larger $N_{\max}$ will be included in the training set, convergence of the radius extrapolations would not be achieved for the unbound excited state $(3^{+},0)$ of $^{6}$Li, which decays through the ${\alpha+d}$ channel.

The ability of our extrapolation technique to capture trends in extended observables like rms radii suggests that the method can also be successfully applied to other observables like, e.g.,transition strengths.
Because the ensemble approach provides intrinsic error estimates, this methodology naturally complements existing uncertainty quantification efforts. Overall, our work encourages development of systematic pipelines wherein high-performance many-body calculations feed into neural networks, thereby extending the predictive reach of ab initio nuclear models.

\section*{Acknowledgments}
This work was supported by the Ministry of Science and Higher Education of the Russian Federation (project~\#FEME-2024-0005).

\bibliographystyle{elsarticle-num}
\bibliography{lib.bib}
%\nocite{*}

\end{document}